# ON THE MINIMIZATION OF HANDOVER DECISION INSTABILITY IN WIRELESS LOCAL AREA NETWORKS

Meriem Abid[1,2], Tara Ali Yahiya[3] and Guy Pujolle[1]

[1]Laboratoire d'Informatique de Paris 6, Paris, France
Firstname.Lastname@lip6.fr
[2]Ginkgo Networks, Paris, France
meriem.abid@ginkgo-networks.com
[3]Telecom SudParis, Evry, France
tara.ali-yahiya@u-psud.fr

### ABSTRACT

*This paper addresses handover decision instability which impacts negatively on both user perception and network performances. To this aim, a new technique called The HandOver Decision STAbility Technique (HODSTAT) is proposed for horizontal handover in Wireless Local Area Networks (WLAN) based on IEEE 802.11standard. HODSTAT is based on a hysteresis margin analysis that, combined with a utility-based function, evaluates the need for the handover and determines if the handover is needed or avoided. Indeed, if a Mobile Terminal (MT) only transiently hands over to a better network, the gain from using this new network may be diminished by the handover overhead and short usage duration. The approach that we adopt throughout this article aims at reducing the minimum handover occurrence that leads to the interruption of network connectivity (this is due to the nature of handover in WLAN which is a break before make which causes additional delay and packet loss). To this end, MT rather performs a handover only if the connectivity of the current network is threatened or if the performance of a neighboring network is really better comparing the current one with a hysteresis margin. This hysteresis should make a tradeoff between handover occurrence and the necessity to change the current network of attachment. Our extensive simulation results show that our proposed algorithm outperforms other decision stability approaches for handover decision algorithm.*

### KEYWORDS

*Handover Decision stability, Ping-pong Effect, Autonomic Networks, Piloting Plane Concept*

## 1. INTRODUCTION

With the deployment of a myriad of new technologies such as 3G/UMTS based on 3GPPP, WiFi based on IEEE 802.11, 4G/LTE based on 3GPPP2 and mobile WiMAX based on IEEE 802.16e, future wireless networks will be characterized by their heterogeneity and their complementarity in terms of coverage area, data rate, security and so on. In such context, Mobile Terminals (MTs) have more and more the opportunity to be connected to the most appropriate access network with regard of the service cost, QoS, network performances. Seamless intra and inter system handovers have a centric role to play in guarantying the MTs the attachment to the most appropriate access network.

The handover procedure intends to move from one network to another along with maintaining MTs' connectivity while moving from one attachment point to another one [1]. Usually, the handover procedure is described through the two phases namely the handover decision phase and the handover execution phase. Since there is no decision without information, there is a third phase called information phase. The decision phase is divided into two steps. It stands for deciding, in one hand, whether to change or not the current connection (if specific conditions are verified); this is known as the initiation step. On the other hand, it has to determine the best access network around; this is known as the selection step. During the handover execution phase, the radio link is transferred to the new network. This step covers layer 2 (Medium Access Control) and layer 3 IP (Network layer) processes and may include the authentication, authorization, and transfer of user's context information. The information step stands for collecting and sharing for both decision and execution steps.

The horizontal handover in WLAN networks consists of an association once the Mobile Terminal (MT) is under the coverage of the candidate Access Point (AP). The horizontal handover is a break before make process (or hard handover) i.e. an association to the new AP takes place only when the MT is disconnected from its old AP. 802.11-based networks carries on a handover only when a client's service degrades to a point that connectivity is threatened. The disconnection time induced is longer than what can be tolerated by highly interactive applications such as voice telephony [2]. In addition, the performance of the target AP is not known until the handover takes place. As a consequence, MTs are always connected to the AP offering the strongest signal. However, this parameter solely is not sufficient to guarantee a good quality of service to the users' applications. This is due to the fact that incorrect handover decision impacts directly on the users' service quality inducing handover decision instability.

The ultimate objective of this article is to fix the decision instability of the handover decision approach adopted in the first version of our handover piloting system [3]. The effectiveness of the handover piloting system has been evaluated on a Linux-based experimental test bed, but there was no mechanism that addresses the handover decision stability issue. To this end, we propose HandOver Decision STAbility Technique (HODSTAT) that specifically addresses the decision stability problem in horizontal handover caused mainly by the lack of information and poor decision mechanisms in WLAN networks.

Traditionally, decision instability and ping-pong effect refer to the same phenomenon. However, in this work, we assume that the decision instability encompasses the ping-pong effect. Indeed, decision instability characterizes unnecessary and inefficient handover decisions. The ping-pong effect is a specific case that is expressed by an important number of comings and goings of the MT in *a short period of time*. The decision instability often leads to performance worsening by causing an interruption of the service.

To minimize decision instability in the handover process, the approach that we advocate should reduce unnecessary handover occurrence while maximizing at the same time the connection quality.

The rest of the article is organized as follows. The next section presents a state of the art that details several techniques that tackle handover decision instability problem. Section 3 summarizes the main components of the handover piloting system. In section 4, we describe The Handover Decision Stability Technique. Section 5 is dedicated to the evaluation performance. Finally, we conclude this work with a conclusion and some perspectives.

## 2. RELATED WORK

Extensive work in the literature considered the ping-pong effect pointing out to the decision instability in the handover decision phase. Indeed, throughout our work, we address more generally the problem of handover necessity especially in the case when utility function for handover decision is used. In this section, we discuss the different techniques that are proposed in the literature to address the ping-pong effect and the handover decision instability as a whole. Usually, the ping-pong effect designates the multiple handover decisions that are triggered in a short period of time. The ping-pong effect usually occurs in the edge of cells while the signal strength of two adjacent access networks is relatively close. Thus, this phenomenon is especially observed in the case of classical decision approaches that are based only on the signal strength or related parameters.

In order to avoid the ping-pong effect in horizontal handover, different techniques such as threshold, hysteresis, dwelling timer or waiting time can be used either alone or jointly in the handover decision process [4]. These techniques can also be adopted to address the handover decision instability.

In [5] a hysteresis is defined to improve the performance of Mobile IP. A handover is executed if the signal strength of the new attachment point is greater than the one of the old attachment point by a fixed hysteresis margin. In [6], the authors propose an adaptive handover algorithm for cellular networks with a dynamic hysteresis value, based on the distance between the mobile station and the serving base station. They compare their proposal to a fixed hysteresis algorithm, both hysteresis and threshold algorithm and an algorithm based on the distance combined with hysteresis. Other advanced techniques are also provided in [7] where a variable hysteresis is adopted to avoid unnecessary handovers. The authors in [8] and [9] introduce a hysteresis cycle or a waiting time in the handover decision. This waiting time is defined as the minimum interval of time that should be assured between two consecutive handovers.

The handover decision stability can also be considered regarding the synchronization problem [10] in addition to the ping-pong effect. The synchronization problem is typical when performing mobile controlled handover (MCHO). Indeed, MTs detect essentially at the same time that more than one access network is offering good performance and decide to hand over at the same time. A good solution is proposed in [10], which consists in introducing randomization into the decision mechanism. The authors proposed a stability interval time which is defined as the waiting period before handover. A random number is generated, as the waiting period before handovers. The problem is then solved with a randomized stability period.

More mechanisms are proposed to avoid unnecessary handover. However, these different approaches are mainly applied in the case of signal-based handover algorithms [6] [11] which may lead to bad decision. Indeed, the degradation of the signal level is a random process especially in case of WLAN and techniques that address the handover stability for algorithms based on signal strength measurements may not be successful to avoid the ping-pong effect [4].

In HODSTAT, we used a hysteresis margin in combination with a utility-based function that take into account different and sometimes contradictory objectives. Our approach ensures a good decision stability comparing to other approaches that address only the ping pong effect.

Approaches that address handover decision stability may result in degradation of QoS and even break up the current application if the handover is delayed. Therefore, there should be a tradeoff in the handover decision algorithm between handover necessity and the number of handover execution while choosing the hysteresis value.

## 3. GINKGO FRAMEWORK FOR AUTONOMIC HANDOVER PILOTING

Our application was built upon a software platform developed by Ginkgo-Networks and called "Ginkgo Piloting Agent" [12]. The purpose of this platform is to facilitate the development of network applications that aim to increase autonomicity in the network.

In our approach, we mainly focus on the decision and the information steps. The handover piloting system should deal with challenges posed by the multi objectives optimization problem. Therefore, it relies on a novel approach based on the use of autonomic principles.

Using the handover piloting system in WLAN allows MTs to be continuously connected to the AP that is offering the best performance in term of QoS and regarding both user/operator preferences and applications' requirements.

Given the current status and practical limitations of WLAN based on IEEE 802.11, MTs cannot rely on any feedback from the APs [13]. The information recovery is so critical in these networks that almost all handover decision metrics are related to the signal strength, the only parameter that enables to compare different access networks.

The handover piloting system is distributed over the network due to the use of Ginkgo Agent technology. In the following sections, we describe the Ginkgo framework and its entities for piloting handover decision.

### 3.1. Piloting Agents' deployment

The autonomic handover piloting system is based on the Ginkgo Piloting Agent (GPA) technology [12]. The GPA is running on different devices and requires few modifications in the Open Systems Interconnection (OSI) protocol stack in order to bypass the current implemented algorithms.

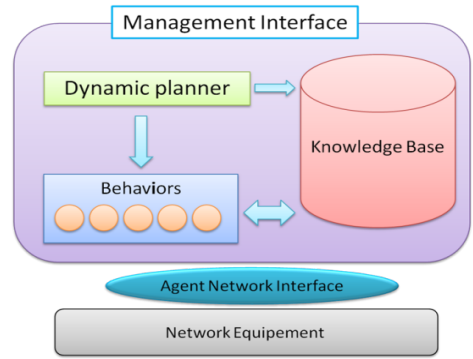

Figure 1. Piloting Agent architecture

As pictured in Figure.1, each GPA is composed of a knowledge base where the knowledge is stored, the whole knowledge of an agent related to its own equipment and its environment (neighboring agents) is represented there. Using the knowledge from the Knowledge Base, the GPA executes periodically a sequence of processing units called Behaviors. Each Behavior is responsible for a part of the global operation of the GPA. The dynamic planner is a specific scheduler that controls and orchestrates the work of Behaviors. The Management Interface gives the operator access and control over the whole knowledge in the Knowledge Base, and some control parameters in the Behaviors and Dynamic Planners, especially by uploading new Policies to the agent's Dynamic Planner. The Agent Network Interface enables GPA to retrieve data/measurement from the network equipment and at the same time to act upon it for piloting a dedicated application.

In the proposed piloting system, we distinguish different sorts of GPA depending on their functions/responsibilities. In the context of WLAN, we deploy two sorts of GPA:

• MT agent which is embedded in the MT equipment. It is responsible for initiating the handover and making the appropriate decisions.

• AP agent which is embedded in the IEEE 802.11 APs. It is responsible for collecting and sharing information.

### 3.2. Handover Information Process

The handover information process addresses both the following aspects: (1) the handover decision metrics that the GPA manages in one hand and (2) the information diffusion configuration on the other hand.

#### 3.2.1. The Handover Decision Metrics

Assessing the quality of current and target access networks requires good predictors from which can be derived meaningful knowledge.

Traditional criteria used for the handover initiation and selection implemented in WLAN are almost all based on the RSS (Received Signal Strength) or signal related parameters e.g. SNR (Signal to Noise Ratio) SIR (Signal-to-Inference Ratio), CIR (Carrier to Inference Ratio) [14]. However, temporary fluctuations of the signal due to different environment conditions (like channel condition) and also to network card may lead to unnecessary handover.

Many works [15], [16] highlight the need to use more eloquent parameters. Unfortunately, most of them are usually hard to retrieve in real environments. In our solution, we choose handover criterions related to the physical and Media Access Control (MAC) layer parameters. We envisage integrating in the forthcoming work more decision metrics such as parameters from other layers (network or application, etc.). Other criterions related to the user preferences/needs will be taken into account as well. The information involved in the decision step comes from the knowledge plane.
Collecting useful information is also critical and should result from a good compromise between information update frequency and network overload. This is controlled through the use of information exchange mechanism of neighboring agents.

### 3.2.2. Information Diffusion Configuration

One of the main critical issues in the wireless technologies that affect seriously on the handover decision process is the physical inability of terminals to communicate with "the unassociated" surrounding APs. Therefore, terminals cannot retrieve any performance information which can allow comparisons and decisions. To overcome this shortage, we set up an agent communication protocol neighborhood-based that makes each AP agent sending updated information to its neighbor agents frequently. This mechanism offers an efficient retrieval of external information. Indeed, each AP agent's Knowledge Base sends performances information periodically to its neighbor agents. Hence, each terminal becomes able to get information from all "unassociated" (but sensed) APs via its associated AP. This facilitates performing the handover decision stage.

The communication of agent protocol is pictured in Fig.2 and can be summarized as follows:
- AP agents send their knowledge to all other AP agents at one hop in the wired backbone network and to all the Terminal agents of the associated terminals.
- Terminal agents receive their knowledge to the AP agent of their associated AP.

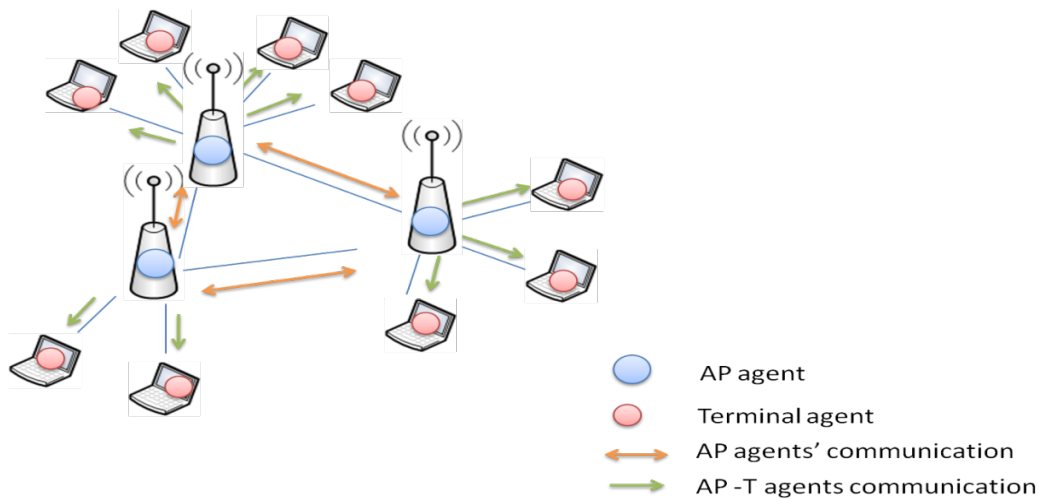

Figure 2. Agents' communication protocol

### 3.3. Handover decision

Generally, decision strategy in WLAN is usually the same for initiation and selection processes. Traditionally, it is based on signal strength related parameters [14]. Our handover decision approach is not based on the signal strength only. We define different decision criterions that assist the utility function to trigger the handover when necessary and at the appropriate time.

### 3.3.1. Utility-based handover function

Network decision process is relevant for maintaining a good performance quality to MTs. In the handover piloting system we used a utility-based function to initiate a handover and select the best access network around.

In decision theory, utility functions can be used to model agent's behavior in situation of choice. Utility functions are studied especially in microeconomics theory. It is used here to represent the agent's preferences. Some properties of the utility function $U$ are:
-   $U$ should increase strictly monotonically when a parameter $x_i$ from the vector $x$ increases and other parameters are kept constant.
-   The decision vector x is containing all "good" parameters defined as decision criterions (i.e the greater value is the better); e.g. we take the bandwidth directly but 1/delay should be used instead of delay.

The utility function that we used is $U(x) = 1 - e^{-\alpha x}$, where $\alpha$ is the relative weight of the decision vector $x$. To get a final handover decision, we used a score combination as shown bellow.

First, a score is computed for each protagonist's objective (user's, application's, ...), as introduced in [17]. For a given user $u$, the score $S_u^n(a)$ is evaluated for a given network $n$ and for each objective $a$.

$$S_u^n(a) = \sum_{x=1}^{k} 1 - e^{-\alpha x} \qquad (1)$$

Where $x$ is the decision vector and $\alpha$ its associated weight.

For the application score (i.e. $a$ is the application's objective), $x$ are the performance parameters such as bandwidth, delay, error, etc. It is worth noticing that application's score value is reduced to zero if the QoS performances offered by the associated network are less than the QoS required by the application.

These scores are then combined using the weighted sum method. Let $C_u^n$ the score combination for the different objectives computed for a given network $n$.

$$C_u^n = \sum_{a=1}^{q} w_a S_u^n(a) \qquad (2)$$

Where $a$ is the protagonist's objective and $w_a$ is the associated weight.
$N$ denotes the set of the candidate networks $c$ sensed by the user $u$, $F_u$ is the set of the score combination values of the candidate networks such as $F_u = \{C_u^c, \forall c \in N\}$.
$C_u^{c\_best}$ is the combination score of the best candidate network and is obtained by

$$C_u^{c\_best} = \sup(F_u) \qquad (3)$$

The handover decision consists then, in a comparison between the combination score of the associated network $C_u^{asso}$ and the best candidate network $C_u^{c\_best}$. The handover is initiated or a new target network is selected if the best candidate network results in the highest computed combination score value

$$C_u^{c\_best} > C_u^{asso} \qquad (4)$$

## 4. HODSTAT: THE HANDOVER DECISION STABILITY TECHNIQUE

The utilization of HODSTAT in combination with the utility-based function (located in the Handover decision) aims at enabling MTs to determine whether or not initiating a handover when the handover should be initiated and which AP will be the target automatically without any user interaction.

In this section we describe the novel handover decision algorithm in the piloting system with the new integrated HODSTAT module.

## 4.1. HODSTAT description

The decision stability is an important feature that should be part of every good handover decision function.

The proposed utility-based handover decision algorithm may be computed at user or network side. In our approach, we adopted a mobile controlled MCHO and network assisted (NAHO) handover strategy using the agent communication protocol. The utility-based handover algorithm is thus implemented as a dedicated behavior performed by MT agents.

Periodically, MT agents compute the combination score $C$ of both associated and candidate networks thanks to information sent by the associated AP agents. Then, each MT agent compares combination scores of the best candidate and the associated network.

To reduce unnecessary handover occurrences, the handover decision is done using a Hysteresis margin. Indeed, MT agent initiates a handover (or selects the best candidate access network) if the score of the best candidate network is greater than the one of the associated network with a Hysteresis margin $H$ i.e. :

$$C_u^{c\_best} > C_u^{asso} + H \qquad (5)$$

MT remains associated to the same network otherwise.

The value of the Hysteresis margin plays a key role in handover performance and affects network performance. We have conducted some experimentation in order to find a good compromise between the handover occurrence frequency and the QoS performance of the associated access network.

## 5. PERFORMANCE EVALUATION: THE HANDOVER ASSESSMENT

We used computer simulations to evaluate our proposed stabilization decision method. We started by modeling WLAN-based wireless environment as well as the user mobility model within the simulated environment. In order to study the performance of our approach, we compare it with the well known "waiting time approach" [8].

## 5.1. Simulation and Environment Model

We have developed a simulator in Java 1.5 [18] that integrates the ginkgo agent technology[1] and implements wireless network environment. The developed simulator is mono-process and runs in a discrete time. In addition, it allows recovering the following information:
- Network topology and its changes (e.g. due to node mobility)
- The flows
- Network performance metrics
- User preferences

### 5.1.1. The Mobility Model: Random Way Point

User mobility trajectories are characterized by the widely used Random Way Point (RWP) model.

At the beginning, each MT begins by staying in one location for a certain period of time (i.e., a pause time).

Once this time expires, the MT chooses uniformly a random destination point (or waypoint) in the simulation area. The user moves to this destination with a speed that is

---

[1] The ginkgo agent acts as in a real environment

International Journal of Computer Networks & Communications (IJCNC), Vol.2, No.3, May 2010

uniformly distributed between [minspeed, maxspeed] (these two values are the same in our simulation = 0.8 m/s).

The MT then moves toward the newly chosen destination at the selected speed.

Upon arrival, the MT pauses for a specified time period (uniformly distributed between [minpause, maxpause]) before starting the process again.

### 5.1.2. Simulation Parameters

We have conducted our simulations considering the following network conditions. The simulation configuration is summarized in the table 1:

| Simulation parameter | Value |
|---|---|
| Simulation time | 75 seconds |
| Mobility model | Random Way Point |
| Mobile velocity | 0.8 meter/second |
| Number of user | 52 |
| Number of MTs | 14 |
| Mobility users ratio | 27% |
| HO_decision_behaviorTimeStep | 0.5 seconds |

Table 1. Simulation parameters setting

### 5.1.3. The Evaluation Criterions

We assess the proposed stabilization decision approach by defining the following evaluation criterions namely the HandOver rate (HO rate) and the corresponding score value (Score rate).

HO rate reflects the number of handovers that occur during the simulation for each MT.

The Score rate is the sum of the score value of the associated access network computed per HO decision behavior time step for each user during one simulation. This evaluation criterion describes in term of score value the performance of the associated access network.

$$HO\_rate = \frac{1}{nb\_MT} \sum_{i=1}^{nb\_MT} nb\_HO_i \qquad (6)$$

$$Score\_rate = \frac{1}{nb\_MT} \sum_{u=1}^{nb\_MT} (\frac{1}{nb\_Step} \sum_{1}^{nb\_Step} C_u^{n\_asso}) \qquad (7)$$

The number of step $nb\_Step$ is calculated as shown in equation (8):

$$nb\_Step = \frac{simuTime}{HO\_decision\_behaviorTimeStep} \qquad (8)$$

These evaluation criterions are used for two following experimentations. The first one consists in choosing a good value to the Hysteresis margin while the second one is set up for comparing our approach with the waiting time approach.

### 5.2. Assigning the Hysteresis margin value:

The rationale behind using a hysteresis-based approach is to avoid handovers when the performance of the candidate access network is almost similar to the associated access network. In such situation, our approach will keep the MT associated to the current access point which reduces disconnection time that only causes performance degradation and unnecessary costs. Indeed, considering that each AP switch involves time consuming authentication and re-association procedures (e.g., see [19]). Hence, it is important to avoid unnecessary handovers by carefully assigning the appropriate Hysteresis margin value.

This value should be chosen to reduce unnecessary handovers by remaining associated to the current network when its performance is close to the candidate networks. However, the Hysteresis margin value should not be too high in order to detect quickly performance collapse in the current associated access network.

### 5.2.1. Simulation Results

To choose the best Hysteresis margin value, 10 simulations were conducted for each of the 21 proposed Hysteresis values. The simulations include 14 MTs which represents 140 experimentations per hysteresis value.

In this article, we have considered that the best Hysteresis margin value is the one that minimizes the handover occurrence while maximizing at the same time the score rate i.e. Min HO_rate and Max Score_rate.

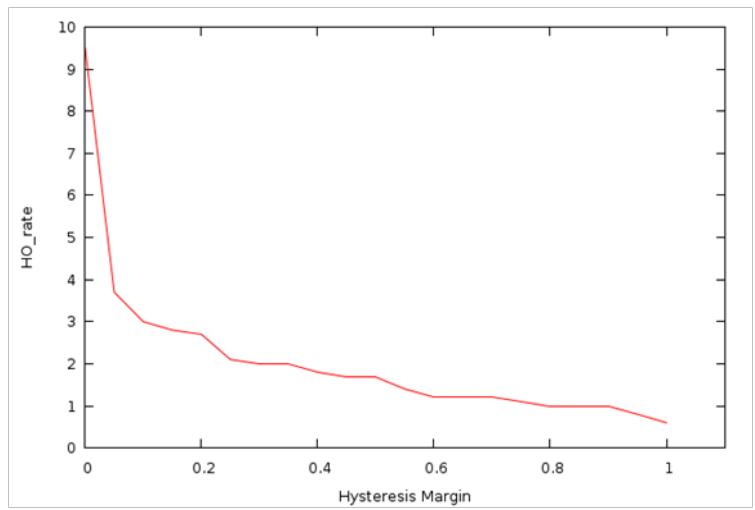

Figure 3. Hysteresis Margin - HO rate

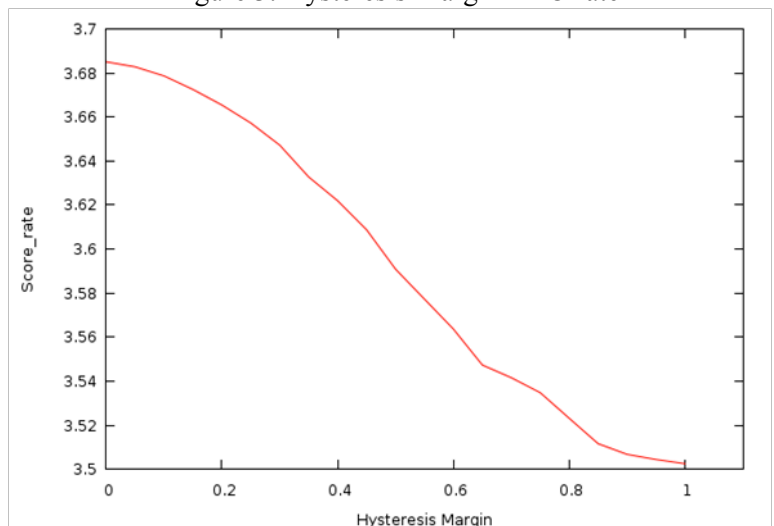

Figure 4. Hysteresis Margin - Score rate

We have chosen to show for each Hysteresis value (21 values) the worst cases i.e. the maximum handover occurrence (Figure.3) with their corresponding score values (Figure.4). The best Hysteresis value is the one that minimizes the handover occurrence in the worst case and maintains a good score rate at the same time. As we can see in Fig 3 and Fig 4, the appropriate value of Hysteresis is in the interval [0.45, 0.55].

### 5.3. Comparing HODSTAT to the waiting time approach

We have conducted some tests that compare our approach to the largely adopted waiting time one.
The two solutions are compared using the aforementioned evaluation criterions. A good handover decision should minimize the handover rate while maximizing the score rate.

#### 5.3.1. Simulation Results

Fig 5 and Fig 6 illustrate the HO rate criterion variation for both hysteresis margin and waiting time approaches during the total simulation time for 14 MTs with a confidence interval that is set to 95%.

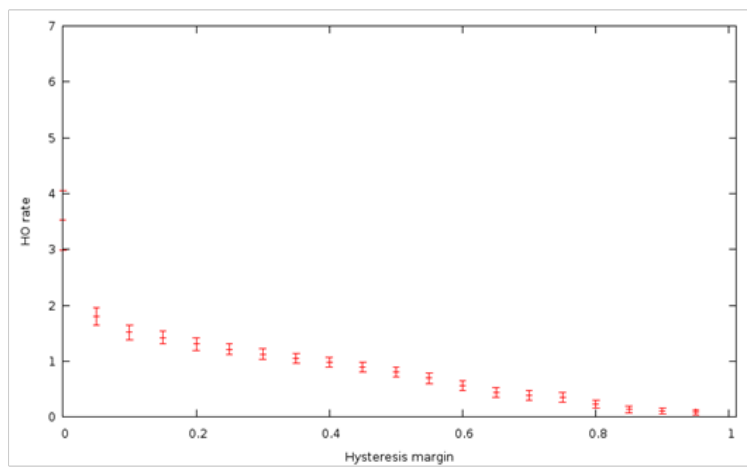

Figure 5. Hysteresis Margin – HO rate

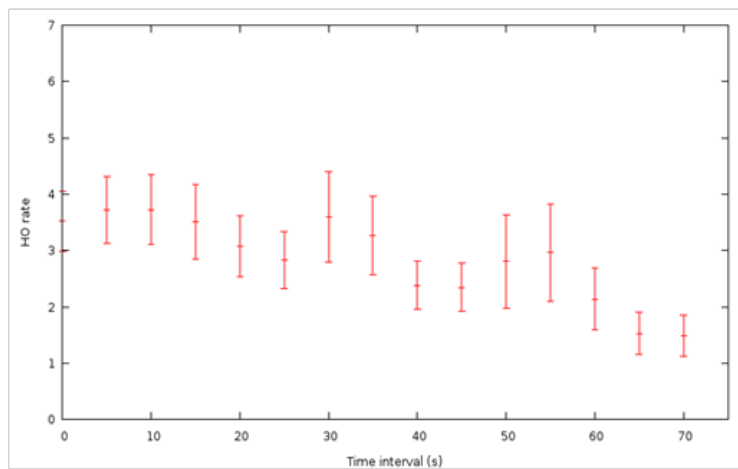

Figure 6. Waiting Time – HO rate

The "0" value for both hysteresis margin and time interval represents the handover decision as implemented in the handover piloting system without any technique to address the handover instability. As we can see, this value implies the highest HO rate observed during a simulation.

We also notice that the HO rate is higher on average in the waiting time approach to the one observed for the hysteresis margin.

In the waiting time approach the HO rate varies considerably from one terminal to another one that is why the confidence interval is so high. This is due to mobile terminals positions regarding the networks' overlapping.

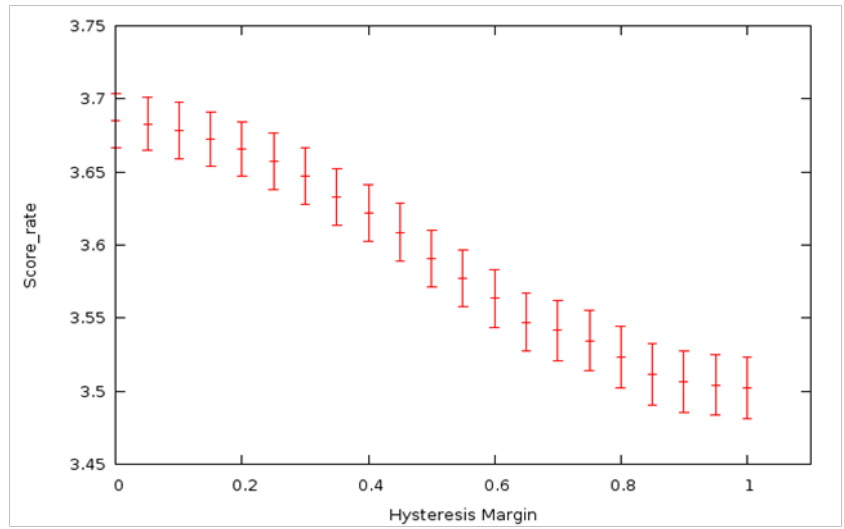

Figure.7 Hysteresis Margin – Score rate

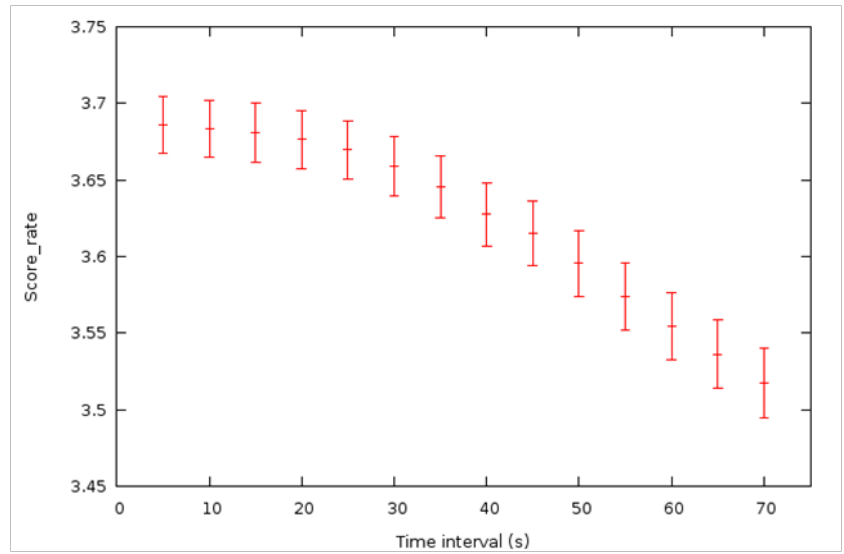

Figure.8 Waiting time – Score rate

In Fig7 and Fig 8, we can observe that when hysteresis margin or waiting time increases, the Score rate diminishes. This is explained by the remaining of MTs associated to the same network despite the existence of better adjacent candidate networks in term of Score rate.

We can also see that both approaches maintain the Score rate relatively high while the hysteresis margin and the waiting time increase.

48

## 3. CONCLUSION AND PERSPECTIVES

In this paper, we have proposed a new approach that stabilizes a utility-based handover decision algorithm in a context of WLAN horizontal handovers. The underlying objective is to eliminate or at least reduce the triggering of handovers that result in performance degradation and wasting of resources due to the user's short visit to the discovered network's coverage area.

Indeed, we advocate that it is better for MT to get a connection with relative QoS for a long time than causing several disconnections for a negligible gain.

Our approach relies on a fixed threshold that determines the suitability of a handover.

To help stabilizing the utility-based function that we utilized in the handover piloting system, we proposed to study the best value of the hysteresis margin between the utility value of the associated and the candidate access network. In addition, we compared our approach to the classical solutions that define a waiting time after handover occurrence. The simulation results show better performance for HODSTAT.

As a future work, we expect to study the performance of HODSTAT in a context of vertical handovers. In addition other aspects of the piloting systems such as the agent communication protocols will be studied in a near future.

**Authors**

**Meriem Abid** received her Engineer Diploma from University of Es Senia Oran in Algeria in 2007. She is currently a PhD student at the Laboratoire d'Informatique de Paris 6 (LIP6), Paris France and a research engineer in Ginkgo Networks Society, Paris France. Her research interests include wireless and autonomic networks, mobility management in next-generation heterogeneous networks and artificial intelligence.

**Tara Ali-Yahiya** received her M.S. in Computer Science from Gaspard-Monge Institute in the University of Marne-la-vallée in 2004. She received her Ph.D. degree in network and computer science from the University of Paris VI, Paris, France, in 2008. She was a post doctoral fellow in Telecom SudParis, Evry, France, in 2009. She is currently an associate professor in the University Paris Sud 11—Orsay. Her research interests include wireless networks, resource allocation under QoS, network planning and modeling, as well as performance evaluation

**Guy Pujolle** received the Ph.D. and "Thèse d'Etat" degrees in Computer Science from the University of Paris IX and Paris XI on 1975 and 1978 respectively. He is currently a Professor at the Pierre et Marie Curie University (Paris 6) and a member of the Scientific Advisory Board of Orange/France Telecom. He spent the period 1994-2000 as Professor and Head of the computer science department of Versailles University. He was also Professor and Head of the MASI Laboratory (Pierre et Marie Curie University), 1981-1993, Professor at ENST (Ecole Nationale Supérieure des Télécommunications), 1979-1981, and member of the scientific staff of INRIA, 1974-1979.

Dr. Pujolle is the French representative at the Technical Committee on Networking at IFIP. He is an editor for International Journal of Network Management, WINET, Telecommunication Systems and Editor in Chief of the indexed Journal "Annals of Telecommunications". He was an editor for Computer Networks (until 2000), Operations Research (until 2000), Editor-In-Chief of Networking and Information Systems Journal (until 2000), Ad Hoc Journal and several other journals.